\documentclass[a4paper,11pt]{article}
\pdfoutput=1 

\usepackage{jinstpub} 
                     
\title{\boldmath LArQL: A phenomenological model for treating light and charge generation in liquid argon}

\author[a]{F. Marinho,\note{Corresponding author.}}
\author[b]{L. Paulucci}
\author[c]{D. Totani}
\author[d]{F. Cavanna}
\affiliation[a]{Universidade Federal de São Carlos, Araras, SP, 13604-900, Brazil}
\affiliation[b]{Universidade Federal do ABC, Santo André, SP, 09210-170, Brazil}
\affiliation[c]{University of California Santa Barbara, California, 93106, USA}
\affiliation[d]{Fermi National Accelerator Laboratory, Batavia, IL 60510, USA}
\emailAdd{fmarinho@ufscar.br}

\abstract{Experimental data shows that both ionization charge and scintillation light in LAr depend on the deposited energy density ($dE/dx$) and electric field ($\mathscr{E}$). Moreover, free ionization charge and scintillation light are anticorrelated, complementary at a given ($dE/dx$, $\mathscr{E}$) pair. We present LArQL, a phenomenological model that provides the anticorrelation between light and charge and its dependence on the deposited energy as well as on the electric field applied. It modifies the Birks' charge model considering the contribution from the escape electrons at null and low electric fields, and reconciles with Birks' model prediction at higher fields. Deviations from current Birks' model are observed for LArTPCs operating at low $\mathscr{E}$ and for heavily ionizing particles. 
The LArQL model presents a satisfactory description at $dE/dx$ and field ranges for interacting particles in LArTPCs and fits well the available data. Improvements via data sets compilation and “global” fits are also interesting features of the model.}


\proceeding{LIGHT DETECTION IN NOBLE ELEMENTS (LIDINE 2021)\\
  September 14th - 17th, 2021\\
  University of California San Diego, San Diego, USA}

\begin{document}
\maketitle
\flushbottom

\section{Introduction}
\label{sec:intro}

Scintillation light is typically collected in current liquid argon time projection chambers experiments (LArTPCs) for time of interaction measurement as this quantity is used in combination with the ionization charge signals to determine the drift distance of the electrons to the anode planes, and thus allowing event 3D position determination. Photon detection systems for future massive LAr volume experiments will have better coverage (e.g. DUNE Far Detector \#2 $\sim$10 kt) and thus, in addition to timing, will be enabled to exploit their own calorimetry capabilities from scintillation light detection at levels similar to those provided by the TPC through charge collection \cite{dunetdr}. For that, an improved description of scintillation production mechanism is required taking into consideration finer details regarding the energy deposits per particle in the liquid argon volume. This was noticed in recent scintillation data analysis such as the one performed at ProtoDUNE \cite{protodunepaper}. Another practical feature from such improved description would be the provision of light yield estimates in a broad range of electric field values. This is relevant as it allows comparisons or fits of one single model description to the different scintillation light data sets produced in LAr test chambers available in the literature. Also the use of such model in the simulation of massive LAr experiments can be of importance because scintillation light produced outside the TPC active volume, where the electric field varies and is approximately zero, can still reach the photo-sensors, and therefore should be considered in physics event analysis. 

\begin{figure}[ht]
    \centering
    \includegraphics[width=0.43\textwidth]{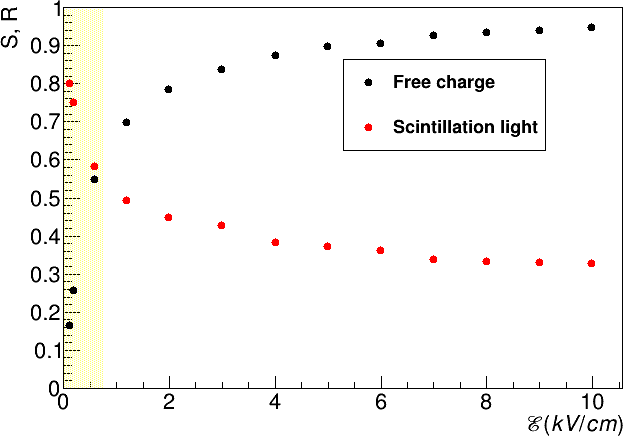}
    \includegraphics[width=0.43\textwidth, height=4.8cm]{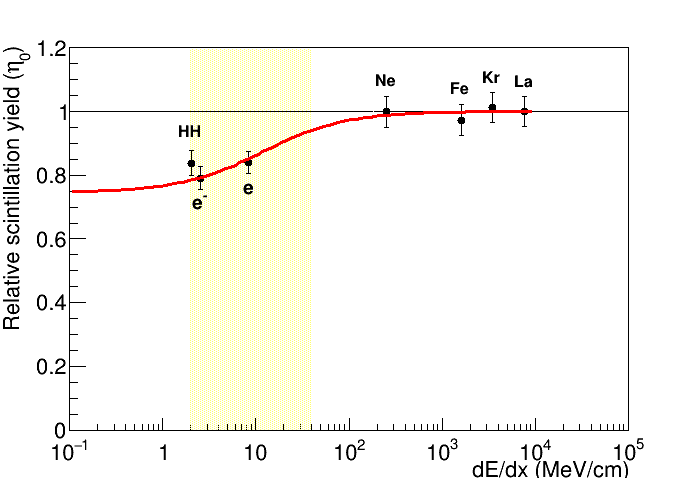}
    \caption{The ratios $S$ and $R$ for {\it mip} in liquid argon as a function of the applied electric field $\mathscr{E}$ (left). Note the complementarity of light and charge. The relative scintillation yield at null field $\eta_0$ in LAr for various incident particles as a function of $dE/dx$ (right). The regions of applicability of LArQL are shown in yellow ($\mathscr{E}$ from 0 to 0.75 kV/cm and $dE/dx$ from 2 to 40 MeV/cm). Data from ref.~\cite{doke2002}.}
    \label{fig:QLdoke}
\end{figure}

Both ionization free charge and scintillation light in LAr depend on the deposited energy density ($dE/dx$) and the electric field ($\mathscr{E}$) \cite{doke1981,kubota1978,doke1988,doke2002}. Figure \ref{fig:QLdoke} (left) shows the ratio $R$, or recombination factor, of the charge produced at a given $\mathscr{E}$ relative to the charge at an infinite electric field regime (maximum collectable charge) and the ratio $S$ of the light yield at a given $\mathscr{E}$ relative to the light yield at zero electric field (maximum light emitted), for minimum ionizing particles ({\it mip}) in a broad range of $\mathscr{E}$. The narrow range of $\mathscr{E}$ values of interest for LArTPC operation is indicated by the yellow band. The observed data trends show that free ionization charge and scintillation light are anticorrelated, complementary at a given ($dE/dx$, $\mathscr{E}$) pair. This suggests that the free charge yield ($Q$) can be related to the scintillation light yield ($L$) through the number of excitons and ionization electrons initially produced per unit of deposited energy. Another relevant aspect observed in the data, shown in figure \ref{fig:QLdoke} (right), is the reduction of the relative scintillation yield at $\mathscr{E}=0$ (usually named $\eta_0$ \cite{doke1988,doke2002}) in the low $dE/dx$ region which is attributed to escape electrons, i.e. those more energetic electrons, escaping the Coulomb influence of the parent ion after initial ionization, that do not promptly recombine into argon excimer states ($\rm Ar_2^{*}$) and thus reduce the amount of de-excitation photon emission. By taking these experimental evidences into consideration we introduce the LArQL model, described in section \ref{model}, that aims to provide adequate $Q$ and $L$ estimates as functions of $dE/dx$ and $\mathscr{E}$. 

\section{LArQL Model}\label{model}

\begin{figure}[ht]
    \centering
    \includegraphics[width=0.55\textwidth]{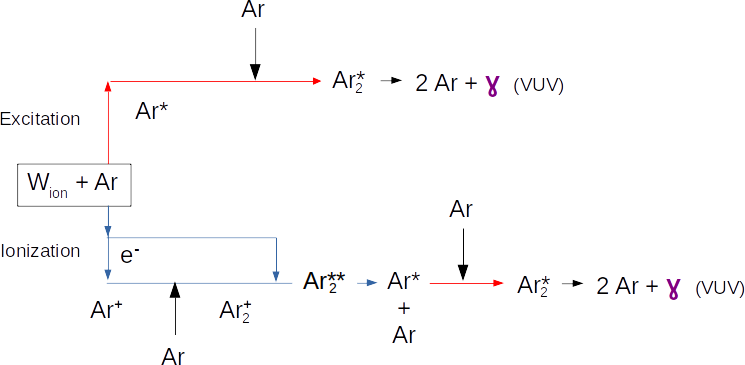}
    \caption{Chain of mechanisms following initial ionization and  scintillation emission pathways in LAr.}
    \label{fig:scint}
\end{figure}

Solid argon is characterized by the existence of an electron band structure and it is usually assumed that the same band structure also exists in the liquid state, characterized by an energy gap of $ E_{gap} = \rm 14.3$ eV. At the passage of charged particles through LAr, besides electron-ion ($\rm Ar^{+}$) pair formation from ionization, Ar* excited atoms are also produced.
The average energy expended per ion pair separation, $W_{ion}$, can be written as:
\begin{equation}
W_{ion} = E_i + \epsilon_{kin} + ( N_{ex} / N_i ) E_{ex}  = 23.6 \pm 0.3 \rm ~eV / e^-,
\end{equation}
where $E_i$ is the average energy loss per ionizing collision corresponding to the mean value of the gap energy and $\epsilon_{kin}$ is the average kinetic energy carried by the sub ionization electron \cite{miyajima1974}. $E_{ex}$ is the average energy release per excited atom and $N_{ex}/N_i = \rm 0.29$ \cite{doke2002} represents the number of excitons formed per electron-ion pair from ionization in LAr. From the above definitions, it follows that the number of ionization electrons initially generated per unit of deposited energy is $N_i=1/W_{ion}$. The chain of mechanisms that can undergo in LAr following initial ionization, recombination and formation of $\rm Ar_2$ excimers and subsequent de-excitation leading to scintillation photon emission, are depicted in figure \ref{fig:scint}. 

A phenomenological model that provides the anticorrelation between light and charge and its dependence with $dE/dx$ and $\mathscr{E}$ is presented here. It aims to cover the range of interest for LArTPC neutrino experiments, $2\,{\rm MeV/cm}<dE/dx<40\rm\,MeV/cm$ 
and $0.25\,{\rm kV/cm}<\mathscr{E}<$ 0.5 kV/cm, and further extends to 0 $< \mathscr{E} <$ 0.75 kV/cm. The model is built upon a well established phenomenological charge-light master equation: 
 \begin{equation}\label{ME}
     Q(dE/dx,\mathscr{E})~+~L(dE/dx,\mathscr{E})~=~N_i~+~N_{ex},
 \end{equation}
$N_i$ and $N_{ex}$ are model input parameters, with current numerical values extracted from data.

LArQL modifies the Birks' charge recombination model correcting for additional escape  electrons freed away from the Ar ions in the lower electric field range, and correlates this with the scintillation light yield through eq. \ref{ME}\footnote{Another commonly adopted model for charge is the Modified Box model \cite{box}, whose free charge output based on a different recombination theory results indeed very close to the Birks' model \cite{icarus2004} output in most of the range of applicability. LArQL implements and corrects this model as well (not reported here), in alternative to Birks' model.}. 
Considered the evidence of light reduction at $\mathscr{E} =0$ for lower $dE/dx$ values, figure \ref{fig:QLdoke} (right), one can establish the fraction of missing photons 1-$\eta_0$ and from this the fraction of escape electrons, $\chi_0$, can be inferred:

\begin{equation}\label{eta0}
    \eta_0 = \frac{1-\chi_0+N_{ex}/N_i}{1+N_{ex}/N_i} ~~~~~~~~~~\longrightarrow ~~~~~~~~~~\chi_0 = (1+N_{ex}/N_i) \cdot (1-\eta_0).
\end{equation}

At any $dE/dx$ at zero electric field, $Q_0 = \chi_0\cdot Q_{\infty} \ne 0$ due to the escape electrons, where the subscripts $0$ and $\infty$ indicate $\mathscr{E}$ value.  Hence, $\chi_0  \cdot Q_{\infty} \leq Q(\mathscr{E}) \leq Q_{\infty}$. However, in the Birks' charge parametrizations, $Q (dE/dx, \mathscr{E} \to 0) \to0,$ which is in friction with the escape electrons evidence. Therefore, a correction term to the Birks' charge model is introduced in LArQL as: 
\begin{equation}\label{qy}
Q_{LArQL} = Q_{Birks} + Q_{ee} = \frac{A_B/W_{ion}}{1 + \frac{k_B}{\rho_{LAr}}\cdot \frac{1}{\mathscr{E}}\cdot \frac{dE}{dx}} + \chi_0\left({dE/dx}\right) \cdot f_{corr}(\mathscr{E},dE/dx)\cdot Q_{\infty},
\end{equation}
where $Q_{LArQL}$ is the charge yield obtained with LArQL, $Q_{Birks}$ is the Birks' model estimate (depending on constants $A_B$, $k_B$, and LAr density, $\rho_{LAr}$) \cite{icarus2004}, $Q_{ee}$ is the escape electrons term, with $f_{corr}$ being the correcting factor accounting for the $\mathscr{E}$ dependence of $Q_{ee}$.

\begin{figure}[ht]
    \centering
    \includegraphics[width=0.43\textwidth]{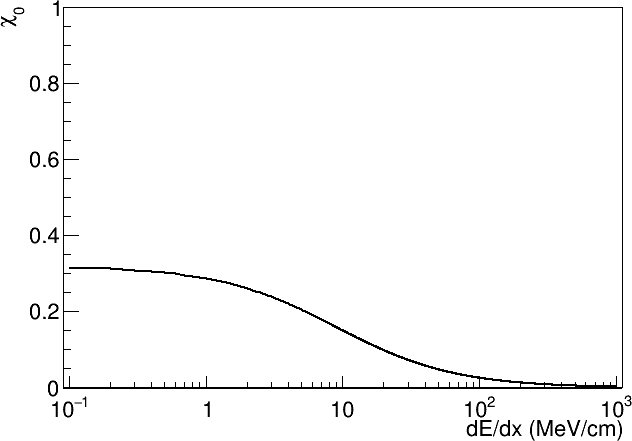}
    \includegraphics[width=0.43\textwidth]{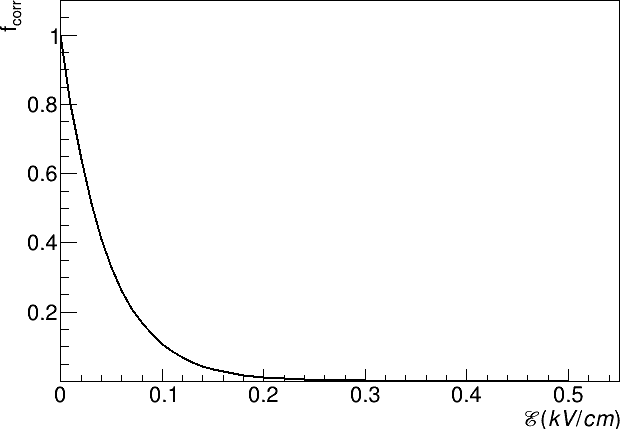}    
    \caption{Running parameters, $\chi_0$ and $f_{corr}$, for the LArQL model. The $f_{corr}$ curve is shown for a {\it mip}.}
    \label{fig:firstfit}
\end{figure}


Based on the experimental data from ref.~\cite{doke2002}, reproduced in figure \ref{fig:QLdoke} (right), the dependence of the $\chi_0$ function on $dE/dx$ was obtained according to eq.~\ref{eta0} and it is shown in figure \ref{fig:firstfit} (left). The $\chi_0$ curve indicates the decrease in the fraction of electrons escaping recombination at $\mathscr{E}=0$ as $dE/dx$ increases. 
As soon as the electric field lifts up above zero, the relative contribution of escape electrons to the charge yield is expected to drop quickly, especially at low $dE/dx$, while the number of electron-ion pairs pulled apart by the electromotive force increases.
A phenomenological function $f_{corr}$ exponentially decreasing with the field is used to parametrize this effect, reducing the $Q_{ee}$ term and restoring the Birks' model ($Q_{LArQL}\rightarrow Q_{Birks}$) at high $\mathscr{E}$, for any $dE/dx$. The $f_{corr}$ curve for {\it mip} in LAr, $dE/dx=2.1$ MeV/cm, is shown in figure \ref{fig:firstfit} (right). The $f_{corr}$ parameter values were also obtained via a LArQL model fit to the scintillation data in ref. \cite{doke2002}.



\begin{figure}[ht]
    \centering
    \includegraphics[width=0.43\textwidth]{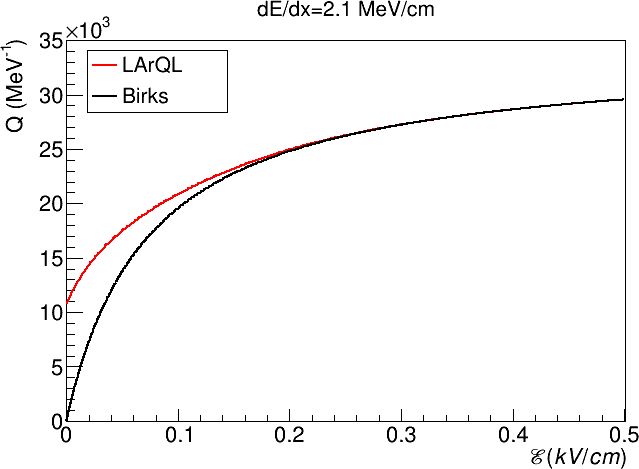}
    \includegraphics[width=0.43\textwidth]{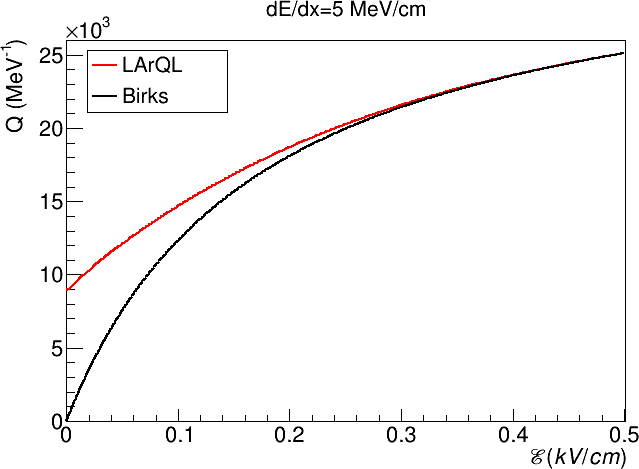}    
    \includegraphics[width=0.43\textwidth]{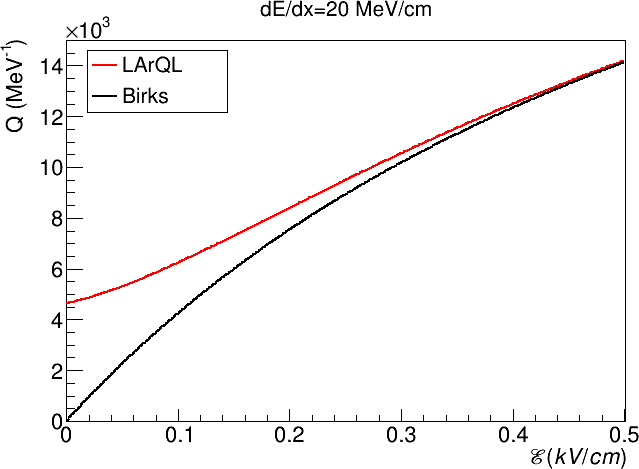}
    \includegraphics[width=0.43\textwidth]{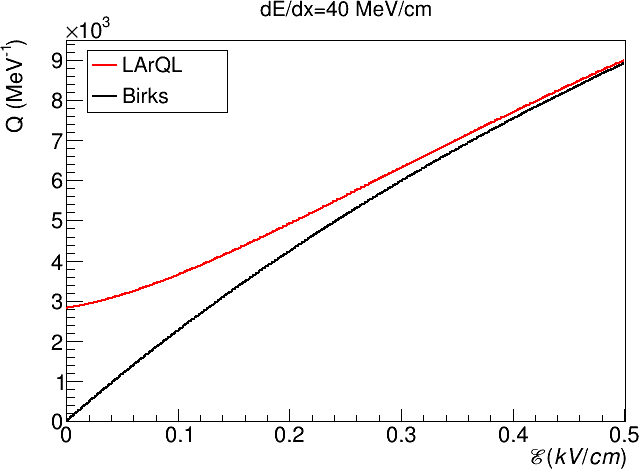}   
    \vspace{-0.2cm}
    \caption{Charge yield Q as function of $\mathscr{E}$ for fixed $dE/dx$: comparison between LArQL and Birks' model.}
    \label{fig:larql_birks_comp}
\end{figure}

Comparisons between LArQL and Birks' model estimates for charge yield as a function of the applied $\mathscr{E}$ for different $dE/dx$ values are shown in figure \ref{fig:larql_birks_comp}. The two models estimates for {\it mip} show appreciable differences only if $\mathscr{E}<0.1 ~\rm\,kV/cm$. As $dE/dx$ increases the difference between the two models gradually becomes more noticeable at higher electric fields ($\mathscr{E} \sim 0.3\rm\,kV/cm$), in the range of current LArTPCs operating values (0.25$<\mathscr{E}<0.5\rm\,kV/cm$). 
The charge yield estimates from LArQL are higher as electrons escaping recombination are taken into account.

Considering that the number of final produced quanta equals the sum of the number of primary ionizations and excitations of argon atoms and that for each recombined electron a scintillation photon is emitted, the light term from LArQL is derived from the charge-light master equation as:
\begin{equation}\label{ly}
L_{LArQL} = N_i - Q_{LArQL} + N_{ex}.
\end{equation}

\begin{figure}[ht]
    \centering
    \includegraphics[width=0.43\textwidth]{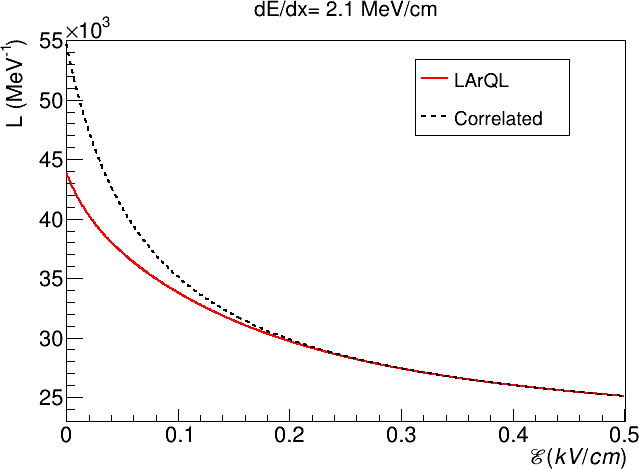}
    \includegraphics[width=0.43\textwidth]{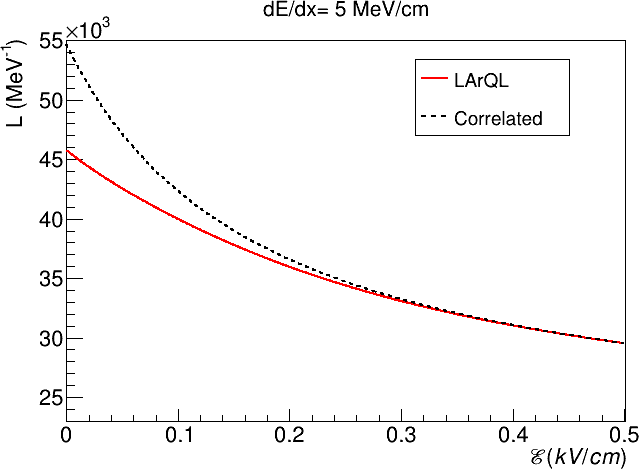}    
    \includegraphics[width=0.43\textwidth]{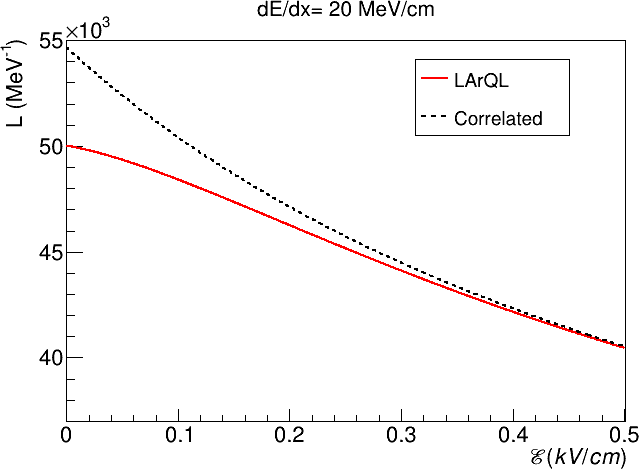}
    \includegraphics[width=0.43\textwidth]{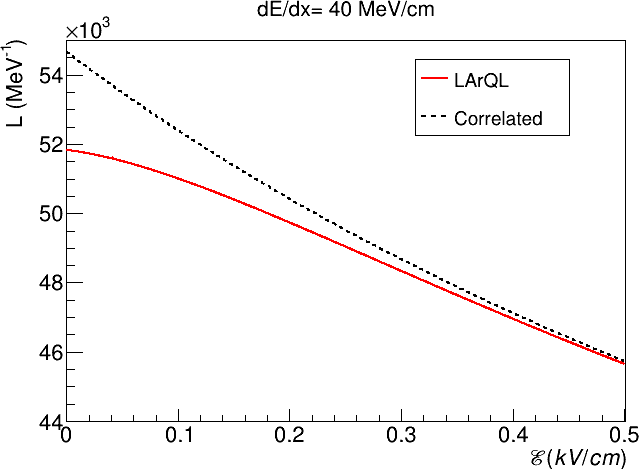}  
    \vspace{-0.2cm}
    \caption{Light yield $L$ as function of $\mathscr{E}$ for fixed $dE/dx$. Red lines indicate calculations from LArQL model and the black dashed lines are estimates from a correlated model not accounting for escape electrons.}
    \label{fig:LY_comp}
\end{figure}

Figure \ref{fig:LY_comp} shows $L$ as a function of the applied $\mathscr{E}$ for different values of $dE/dx$. The decreasing trend obtained for all curves as function of $\mathscr{E}$ is as expected according to the anticorrelation between $L$ and $Q$. The red lines indicate the curves obtained for LArQL with eq. \ref{ly}. The dashed black lines show $L$ for a recently proposed {\it light-charge correlated model} \cite{lariat} which does not take escape electrons into account resulting in the same fixed estimate for $L$ at $\mathscr{E} = 0$ for any $dE/dx$. Notice that LArQL predictions of $L$ differ from the correlated model for any $dE/dx$ at lower values of $\mathscr{E}$ giving estimates for $L$ consistent with available data (e.g. figure \ref{fig:QLdoke}) in all the considered $\mathscr{E}$ range.

\section{First model validation and potentialities}\label{sec:prelim}
\begin{figure}[ht]
    \centering
    \includegraphics[width=0.44\textwidth]{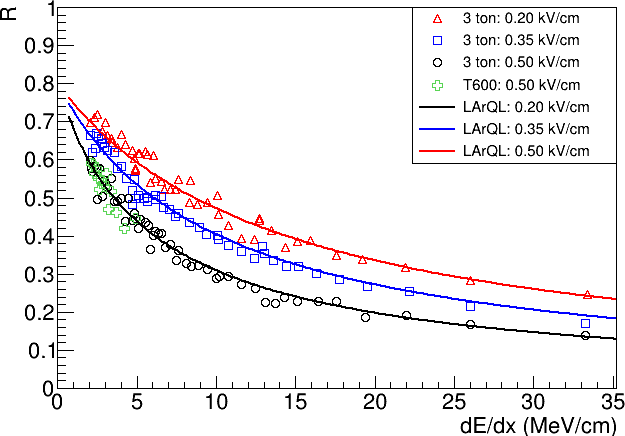}
    \includegraphics[width=0.44\textwidth, height=4.55cm]{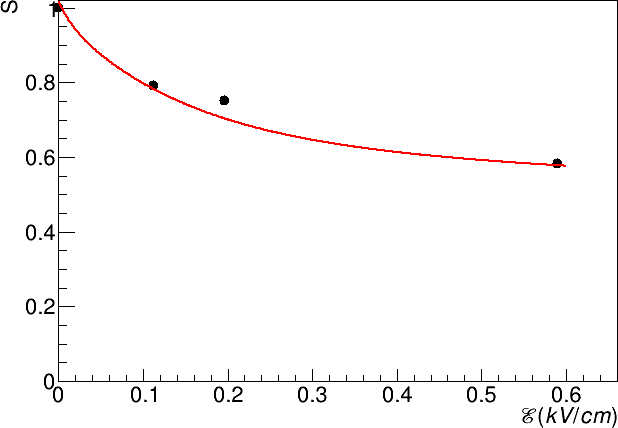}
    \vspace{-0.2cm}
    \caption{Recombination factor $R$ as function of $dE/dx$ for different $\mathscr{E}$ values (left). 
    Data from ref. \cite{icarus2004}. On the right, $S$ ratio for {\it mip} as function of $\mathscr{E}$. Data from ref. \cite{doke2002} for 1 MeV electrons from $^{207}$Bi source in LAr.}
    \label{fig:icarus_r_vs_dedx}
\end{figure}

An initial comparison between LArQL and data is made assuming the corrections described in section \ref{model} (eqs.~\ref{qy} and \ref{ly}) and Birks' model parameters values as from ref.~\cite{icarus2004}. Figure \ref{fig:icarus_r_vs_dedx} (left) shows the charge recombination factor measured in the ICARUS experiment \cite{icarus2004} and the corresponding LArQL estimate. The ratio $S$ between the light detected at a given $\mathscr{E}$ and zero field is also compared with LArQL estimates as is presented in figure 
\ref{fig:icarus_r_vs_dedx} (right). A satisfactory agreement is observed although light data is limited.

Comparison between other measurements and the LArQL model was performed with the same set of fixed parameters. For instance, LArQL provides a better description of the charge density ($dQ/dx$) as function of $dE/dx$ for the data obtained in the MicroBooNE experiment \cite{microboone2020}, with an electric field of 0.273 kV/cm, than the Birks' model. Comparison between light data from the ARIS experiment and LArQL was also performed \cite{aris}. The ratio $S$ as function of $dE/dx$ for different applied electric fields are reasonably described. 

Although LArQL provides a good description of both charge and light data it is reasonable to consider its parameters can still be refined to a more accurate version. A preliminary parameter fitting exercise was performed as a proof of concept for such assumption. ARIS data was also taken in addition to the original aforementioned data. In this, all parameters related to the Birks' formula and $\chi_0$ were kept fixed while the two parameters related to $f_{corr}$ parametrization were allowed to vary in order to minimize model-data residuals. The explicit parametrization for $f_{corr}$ is given by:
\begin{equation}
    f_{corr} = e^{-\mathscr{E}/(\alpha \ln \left(dE/dx\right) + \beta) },
\end{equation}
where $\alpha$ and $\beta$ are free parameters. 

\begin{figure}[ht]
    \centering
    \includegraphics[width=0.49\textwidth]{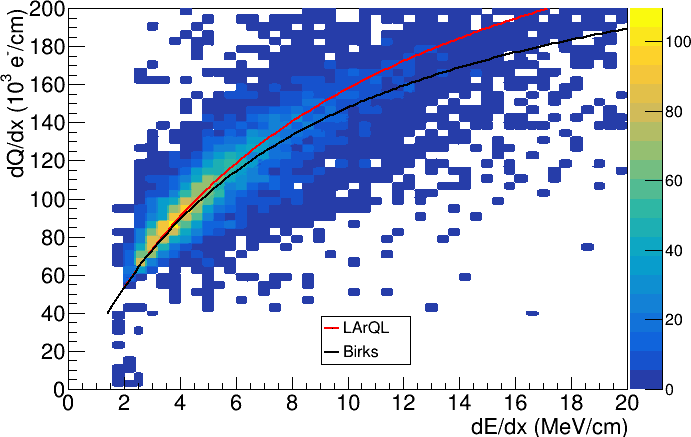}
    \includegraphics[width=0.44\textwidth]{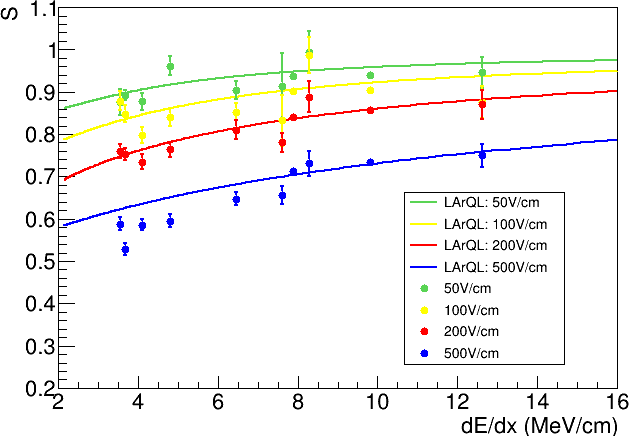}
    \vspace{-0.2cm}
    \caption{Optimized $dQ/dx$ (left) and  ratio $S$ (right) as function of $dE/dx$ comparison. Data from ref. \cite{microboone2020} (left) and \cite{aris} (right).}
    \label{fig:larql_optm}
\end{figure}

By updating the $\alpha$ and $\beta$ parameters in LArQL model one can evaluate its optimization impact on the data description. Figure \ref{fig:larql_optm} shows the improved curves for $dQ/dx$ compared to MicroBooNE data (left) and ratio $S$ compared to ARIS data (right), both as a function of $dE/dx$. The red line in the $dQ/dx$ graph shows LArQL estimates and the black line shows the Birks' model. LArQL better agreement with the charge data is clearer above 5 MeV/cm.
\section{Conclusions}
LArQL is a phenomenological model to evaluate charge and light yields considering that on the passage of a ionizing particle through LAr the final produced quanta are equal to the sum of the number of primary ionizations and excitations of argon atoms. This way it provides the appropriate anticorrelation for the amount of charge and light produced. By considering that electrons can escape recombination at $\mathscr{E}\sim0$ when $dE/dx$ is sufficiently small, it is possible to reproduce data behavior, specially for low $\mathscr{E}$ values. In addition, this model can be potentially improved throughout a global fit where all data available should be considered and more or even all parameters varied. It is worth noticing that given the straightforward model description its direct implementation on the LArSoft platform \cite{larsoft} was possible and it is already available for use.  

\section{Acknowledgment}
This work is partially supported by Conselho Nacional de Desenvolvimento Cient\'\i fico e Tecnol\'ogico (CNPq) under grants 309895/2021-8 (F.M.) and 309891/2021-3 (L.P.). We also acknowledge Funda\c c\~ao de Amparo \`a Pesquisa do Estado de S\~ao Paulo (FAPESP) under Thematic Project 2014/19164-6 (F.M., L.P.).


\begin{thebibliography}{99}
\bibitem{dunetdr}
B. Abi et al. (The DUNE Collaboration),
\emph{Volume IV. The DUNE far detector single-phase technology},
\emph{JINST} {\bf 15} (2020) pg. T08010.
\bibitem{protodunepaper}
B. Abi et al. (The DUNE Collaboration), \emph{First results on ProtoDUNE-SP liquid argon time projection chamber performance from a beam test at the CERN Neutrino Platform}, \emph{JINST} {\bf 15} (2020) pg. P12004.
\bibitem{doke1981}
T. Doke, \emph{Fundamental properties of liquid Argon, Krypton and Xenon as Radiation detector media}, \emph{Portugal Phys.} {\bf 12} (1981) pg. 9.
\bibitem{kubota1978}
S. Kubota, A. Nakamoto, T. Takahashi, T. Hamada, E. Shibamura, M. Miyajima, K. Masuda and T. Doke, \emph{Recombination luminescence in liquid argon and in liquid xenon}, \emph{Phys. Rev. B} {\bf 17} (1978) pg. 2762.
\bibitem{doke1988}
T. Doke, H. J. Crawford, A. Hitachi, J. Kikuchi, P. J. Lindstrom, K. Masuda, E. Shibamura and T. Takahashi, \emph{Let dependence of scintillation yields in liquid argon}, \emph{Nucl. Instrum. \& Methods A} {\bf 269} (1988) pg. 291.
\bibitem{doke2002}
T. Doke, A. Hitachi, J. Kikuchi, K. Masuda, H. Okada and E. Shibamura, \emph{Absolute Scintillation Yields in Liquid Argon and Xenon for Various Particles}, \emph{Jpn. J. Appl. Phys.} {\bf 41} (2002) pg. 1538–1545.
\bibitem{miyajima1974}
M. Miyajima, T. Takahashi, S. Konno, T. Hamada, S.
Kubota, H. Shibamura, and T. Doke, \emph{Average energy expended per ion pair in liquid argon}, \emph{Phys. Rev. A} {\bf 9} (1974) pg. 1438.
\bibitem{box}
J. Thomas, and D. A. Imel, \emph{Recombination of electron-ion pairs in liquid argon and liquid xenon} \emph{Phys. Rev. A} {\bf 36} (1987) pg. 614.
\bibitem{icarus2004}
Amoruso, S. et al. (The ICARUS Collaboration), \emph{Study of electron recombination in liquid argon with the ICARUS TPC}, \emph{Nucl. Instrum. \& Methods A} {\bf 523} (2004) pg. 275–286.
\bibitem{lariat}
W. Foreman et al. (LArIAT Collaboration), \emph{Calorimetry for low-energy electrons using charge and light in liquid argon}, \emph{
Phys. Rev. D} {\bf 101}, (2020) pg. 012010.
\bibitem{microboone2020}
C. Adams et al. (The MicroBooNE Collaboration), \emph{Calibration of the charge and energy loss per unit length of the {MicroBooNE} liquid argon time projection chamber using muons and protons}, \emph{JINST} {\bf 15} (2020) pg. P03022.
\bibitem{aris}
P. Agnes et al. (The ARIS Collaboration), \emph{Measurement of the liquid argon energy response to nuclear and electronic recoils}, \emph{Phys. Rev. D} {\bf 97} (2018) pg. 112005.
\bibitem{larsoft}
E.L. Snider and G. Petrillo,
\emph{LArSoft: toolkit for simulation, reconstruction and analysis of liquid argon TPC neutrino detectors}, \emph{J. Phys.: Conf. Ser.} {\bf 898} (2017) pg. 042057.

\end{thebibliography}
\end{document}